# Carbon riveted Pt/MnO$_2$-Graphene catalyst prepared by in situ carbonized L-ascorbic acid with improved electrocatalytic activity and stability

Hailong Chen

MEMS Center, School of Astronautics, Harbin Institute of Technology, Harbin 150001, China

* Tel: +86 451 86413451; fax: +86 451 86413441; e-mail: asimba@163.com

Abstract

[First Raw Draft]

Pt/Graphene and Pt/MnO$_2$-Graphene catalysts have been prepared by microwave-assisted polyol process. Electrochemical results show that Pt/MnO$_2$-Graphene has higher electrocatalytic activity and stability. Furthermore, carbon riveted Pt/MnO$_2$-Graphene catalyst has been designed and synthesized on the basis of in situ carbonization of L-ascorbic acid. The stability for carbon riveted Pt/MnO$_2$-catalyst is significantly enhanced.

With the development of micro electromechanical system (MEMS), micro direct methanol fuel cell (μDMFC) has been of great research interest due to its simple structure, high energy density and low pollutant emission, and is considered the most potential energy carrier to meet the continuously increasing demand for portable electronic devices [1-8]. Platinum is the best-known precious metal catalyst thatexhibits significant electrocatalytic activity for methanol oxidation [9-12]. However, its low utilization and low stability because of CO poisoning greatly hinder the commercial applications of DMFC [13-17]. In this regard, researchers usually make an alloy of Pt with additional metals or metal oxides (e.g., Pt/CeO2, Pt/Fe3O4, Pt/TiO2, and Pt/SnO2) based on the bifunctional mechanism and the electronic effect [18-20]. Among those widely used metal oxides, MnO2 attracted considerable interest because of its excellent proton conductivity, low cost, natural abundance and environmental friendliness. Some researchers have incorporated MnO$_2$ into commonly used supporting materials, such as carbon black, carbon nanotubes, and the results show that the addition of MnO2 could improve catalytic performance [20-24].

However, due to the relatively low electron conductivity of MnO$_2$ at the cost of catalytic performance, it necessarily deserves the investigation of the structural design of catalyst to weaken the side-effect resulting from the low electron conductivity and the lack of attachment of Pt and MnO$_2$. Recently, some researchers have investigated the MnO2 as a co-catalyst for methanol electrooxidation, but few papers have focused on the structural design of MnO$_2$ based binary catalyst to enhance electron conduction and synergistic effect.

Here we prepare Pt/Graphene and Pt/MnO$_2$-Graphene catalysts by microwave-assisted polyol process. Electrochemical results show that Pt/MnO$_2$-Graphene has higher electrocatalytic activity and stability. Furthermore, carbon riveted Pt/MnO$_2$-Graphene catalyst has been designed and synthesized on the basis of in situ carbonization of L-ascorbic acid. The stability for carbon riveted Pt/MnO$_2$-catalyst is significantly enhanced.

## 2 Experimental

2.1 Synthesis

Firstly, 37.8 mg of MnCl2 and 31.7 mg KMnO4 were separately dissolved in 100 mL of ultrapure water with ultrasonic vibration, stirring each 30 min, then KMnO4 solution was added dropwise to the stirring solution of MnCl2, the mixed aqueous solution was placed in the autoclave and reacted at 200 ℃ for 10 h, then filtered, washed, dried and set aside. Wherein, MnO2 formed were mainly based on the following reaction formula:

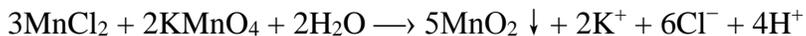
$$3MnCl_2 + 2KMnO_4 + 2H_2O \longrightarrow 5MnO_2 \downarrow + 2K^+ + 6Cl^- + 4H^+$$

Preparation of MnO$_2$/GO: 30 mg GO solid was dissolved in ultrapure water, after that ultrasonic oscillation 30 min, and then 8.6 mg MnCl$_2$ dissolved in ultrapure water, ultrasonic oscillations good dispersion, the MnCl2 solution was added to GO was continued ultrasonic vibration, stirring each 30 min. The KMnO4 solution was added dropwise to a stirring solution of MnCl2, the mixed solution was placed in the reaction at 200 ℃ 24 h, then filtered, washed, dried and set aside.

Preparation of Pt-MnO2/GO: 40 mg MnO2 / rGO solid was dissolved in 10 mL of ethanol, ultrasonic vibration uniformly dispersed. To the above solution was added 40 mL of ethylene glycol, 5 mL glycerol and 5 mL of isopropanol, continued ultrasound, stirring each 30 min. Measure 6.6 mL of H2PtCl6 6H2O ethanol solution (at a concentration of 4mg mL-1) was added dropwise to the above-described MnO2/rGO solution was being stirred, the stirring was continued for 30 min. Adjusting the PH value of the solution was saturated with NaOH in ethylene glycol solution to a value of 12. Remove beaker rotor, and Chapter III except that the heating, the solution was placed in a microwave heater heating interval 10 s, was repeated 6-10 times, so that the temperature rise is 125 ~ 140 ℃. Cooling to room temperature, the solution was adjusted to PH 4 with dilute HNO3 solution. Suction filtered and washed several times with ultrapure water, removing the remnants of Cl-1 plasma. Then placed in a vacuum oven at 60 ℃ 6 h.

Preparation of Pt-MnO2 / GO-L are: 35 mg of MnO2/rGO solid dispersed in ultrapure water, ultrasonic vibration to form a homogeneous slurry, then add 5 mg L- ascorbic acid

(C6H8O6), continues to oscillate evenly dispersed, put it into oven dried, placed in a tube furnace 200 ℃ carbonization, carbonization process using argon gas. Then cooled to room temperature, it will have carbonized coated MnO2/rGO-L catalyst, and then follow the steps for preparing Pt-MnO2 / rGO load Pt, obtain Pt-MnO2 / rGO-L catalyst (PMGL). Fig. 1 shows a flow diagram of the prepared PMGL catalyst.

2.2 Preparation of the working electrode:

The catalyst slurry was prepared by ultrasonically dispersing 4 mg catalyst in the solution of 0.2 ml ethanol and 0.8 ml ultrapure water for 30 min. A glassy carbon electrode (GCE) with the diameter of 4 mm was polished with alumina suspensions and served as the underlying substrate of the working electrode. A quantity of 5 μl of the dispersion was extracted out on the top of the GC followed by drying under room temperature for 4 h.

2.3 Structural and electrochemical measurements:

X-ray diffraction (XRD) and transmission electron microscopy (TEM) were adopted to characterize the morphological and structural properties of the prepared catalyst. The activity and stability of the catalysts were studied by cyclic voltammetry (CV) which was recorded with CHI650D electrochemical analysis instrument with the potential range from xx to xx V and the scan rate of 50 mV s−1. Before recording the measurement curves, the working electrode was treated by continuous cycling at the scan rate of 50 mV s−1 for 50 cycles to get a stable response. To compare the long-term performance of the two catalysts for methanol oxidation, chronoamperometry (CA) tests were used in a solution of 0.5 M H2SO4 and 0.5 M CH3OH for 3600 s with the potential of 0.65V.

**Results and discussions**

Fig. 2 illustrates the XRD patterns for PMGL, PMG, PG and MnO2. For PMGL, PMG, and PG, there are three curves appeared near 25° absorption peak corresponding to amorphous carbon hexagonal graphite characteristic peaks of (002). When viewing peak intensity, PMGL> PG> PMG, indicating that the addition of MnO2 benefits reducing graphene agglomeration, however the generated amorphous carbon during in situ carbonation would then make peak becomes higher. These three catalysts in the absorption peak 39.7 °, 46.1 °, 68.5 °, 80.9 ° and 86.0 ° near the characteristic peak of Pt, corresponding to (111), (200), (220), (311) and (222) plane, indicating that Pt form a face-centered cubic structure. In addition, from XRD patterns for PMGL and PMG, MnO2 catalyst was observed its characteristic pattern, and the position were 37.2 °, 42.8 °, 56.7 °, 66.9 °, corresponding to the (100), (101), (102), (110) crystal plane. According to JCPDS No.30-0820, it can be judged the prepared MnO2 referred to the ε-MnO2 crystalline.

Fig. 3 shows the Raman spectra for PMGL, PMG, and PG. Two graphene Raman peaks in three lines were located 1329 D cm-1 peak and G peak of 1597 cm-1 . Sp2 D band is related to hybridized carbon and defects and disorder, while sp2 G band associated with E2g vibration , which can be used to explain the degree of graphitization. In addition, peak intensity ratio of D and G is relevant with graphene layers and dimensions. after adding MnO2 Pt and the ratio increases, indicating that the phenomenon of graphene stacked get mitigation, defects the active site increases at 630 cm-1 can be observed Raman peaks MnO2, which should be attributed to MnO6 group of Mn-O symmetric stretching vibration.

Fig. 4 shows the TEM photograph and the particle size distribution of PMGL, PMG and PG catalyst. The particle size distribution is obtained by randomly selecting 200 Pt nanoparticles. The Pt particle size for PMGL, PMG and PG particle size was 2.31 nm, 2.34 nm and 2.44 nm, respectively. This shows that the addition of MnO2 to a certain extent reduced the diameter of the Pt particles, and L- ascorbic acid could anchor Pt particles making its size slightly reduced. The dispersion improvement indicated that MnO2 effectively hindered reunion of Pt particles. In addition, from the embedded HRTEM figure for PMGL and PMG, the Pt (1 1 1) crystal plane spacing of 0.22 nm, MnO2 (1 0 0) crystal plane spacing of 0.24 nm, thereby confirmed the as described synthesis process prepared nanograde MnO2 particles and Pt particles, these two particles contact each other, which is likely to alter the structure of each electron cloud.

Fig. 5 presents the XPS spectra for PMGL, PMG and PG catalyst. It can be seen that the three of Pt 4f7 / 2 orbital peak positions were 71.48,71.51 and 72.05eV, and the addition of MnO2 caused a about 0.5 eV negative offset, indicating that there exists strongly interacting electrons between Pt and MnO2 and affect the adsorption on Pt surface intermediate state, that it is possible to make Pt and CO adsorption intermediate product can be reduced, thereby enhancing the Pt resistance to poisoning ability. In addition, the transfer of electrons from Pt MnO2 surface, it is possible to change the state of Mn in the compound, so that the oxide containing Mn more oxygen vacancies, MnO2 ability to enhance the release of the oxygen-containing group, such that CO is more easily oxidized.

Fig. 6 shows cyclic voltammetry of PMGL, PMG and PG catalysts in 0.5 M H2SO4 aqueous solution at room temperature 25 ℃. As can be seen from the figure, PMGL, PMG and PG catalyst for hydrogen desorption peaks successively reduced with electrochemically active area of 52.16 m2g-1,48.92 m2g-1 and 39.27 m2g -1.

Fig. 7 PMGL, PMG and PG catalyst cyclic voltammetry in 0.5 M H2SO4 +0.5 M CH3OH aqueous solution. For positive peaks, in particular, PMGL maximum peak current, of 23.21 mA cm-2, slightly higher than the PMG, which is due on the one hand, although the carbon source can be anchored Pt and MnO2 particles, and its size is reduced to some extent will help enhance the catalytic activity, but on the other hand

around the carbon coated Pt particles will inevitably reduce its active position of the surface is not conducive to enhance the catalytic activity, so the oxidation of methanol curve both comprehensive Results. The results show that the catalytic ability PGML still the highest, which is consistent with their corresponding ESA test results. In addition, PMGL and PMG potential of methanol oxidation starting PG smaller than that in the negative offset, which means improved catalyst at lower potential began the catalytic oxidation of methanol. More importantly, since the ratio of the forward and reverse peak peaks can be directly measured by resistance to CO poisoning of the catalyst, it can be seen from the table If the three of / Ib values were 1.02,0.91 and 0.89, indicating PMGL have better resistance to poisoning, followed by PMG. This may be because $MnO_2$ as an auxiliary component while reducing the Pt and -COads binding energy, on the one hand to promote the generation of oxygen-containing group -OH, so as to enhance the Pt-$MnO_2$ / rGO catalyst CO endurance. In addition, L- ascorbic acid will Pt, $MnO_2$ and graphene more closely together, give full play to the anchoring effect of in situ coated nanoparticles.

Fig. 8 PMGL, PMG and PG three catalysts in 0.5 M $H_2SO_4$ +0.5 M $CH_3OH$ aqueous solution of chronoamperometry curve catalytic oxidation of methanol, setting potential of 0.65 V. Judging from the trend curve, the three have a certain attenuation at the initial stage, which is due to the rapid accumulation of intermediate products such as CO results. Wherein PG catalyst decay fastest, PMG catalyst secondly, PMGL slowest catalyst decay, indicating that the anti-$MnO_2$ and L- ascorbic acid poisoning respects played a beneficial role. 3600 s test at the end of the current density sizes of 2.97 mA cm 2,1.36 mA cm 2 and 0.84 mA cm 2, PMGL a substantial increase in the current value of explanation which has a more excellent anti-drug capabilities and catalytic stability, indicating that the use of great advantage carbonation coating technology, L- ascorbic acid after carbonization of Pt anchoring effect is very obvious.

Fig. 9 PMGL, PMG and PG three catalysts in 0.5 M $H_2SO_4$ +0.5 M $CH_3OH$ aqueous electrochemical impedance spectroscopy catalytic oxidation of methanol. As can be seen from the figure, PMGL minimum arc catalyst, followed by PMG catalyst, the largest is PG catalyst, indicating that the electrochemical impedance PMGL <PMG <PG. This is due to the interaction of Pt oxide added to $MnO_2$ and reduces the internal resistance, in addition, after the addition of a small amount of ascorbic acid carbonized further reduced impedance, making the charge transfer and mass transfer more smoothly, help to further enhance the conductivity.

Fig. 10 presents polarization curves after the assembly of the battery unit cell. In the assembly of passive direct methanol fuel cell, the anode catalysts were the as prepared catalysts mentioned above, and the cathode catalysts employed commercial Pt/C catalyst, The test was conducted at room temperature with 2 M methanol concentration. As can be seen from the figure, PMGL corresponding maximum power density of 26.31 mW cm-2, followed by the PMG of 21.11 mW cm-2, the smallest for the PG of 16.13 mW cm-2, the open circuit voltage was 0.69V, 0.65V and 0.59V, indicating that the catalytic performance of PMGL is best, followed by PMG and PG. The conclusion drawn from single cell test is consistent with the above electrochemical test.

**References:**

[1] Kulikovsky AA. A model for DMFC cathode impedance: The effect of methanol crossover. Electrochem Commun 2012;24:65-8.
[2] H. Chen, H. Jia, C. Zorman, and P. X.-L. Feng, 'Determination of Elastic Modulus of Silicon Carbide (SiC) Thin Diaphragms via Mode-Dependent Duffing Nonlinear Resonances', Journal of Microelectromechanical Systems, 29 (2020) 783-789.
[3] H. Chen, H. Jia, W. Liao, V. Pasheaei, C. N. Arutt, M.W. McCurdy, C.A. Zorman, R.A. Reed, R.D. Schrimpf, M.L. Alles, P.X.-L. Feng, "Probing Heavy Ion Radiation Effects in Silicon Carbide (SiC) via 3D Integrated Multimode Vibrating Diaphragms", Applied Physics Letters, vol. 114, no. 101901, 2019.
[4] H. Chen, H. Jia, M.W. McCurdy, R.A. Reed, R.D. Schrimpf, M.L. Alles, P. Hung, P. X.-L. Feng, "Probing Ion Radiation Effects in Si Crystal by 3D Integrated Resonating Thin Diaphragm", Government Microcircuit Applications & Critical Technology Conference 2019(GOMAC Tech 2019), Albuquerque, NM, Mar. 25-28 (2019).
[5] X.L. Zhang, Y. Li, H. Chen, Y.F. Zhang, X.W. Liu. A water management system for metal-based micro passive direct methanol fuel cells. Journal of Power Sources, 273 (2015) 375-379.
[6] S. Fang, Z. Ma, H. Chen, Y.F. Zhang, X.W. Liu. An Equivalent Circuit Model of Direct Methanol Fuel Cell for Polarization Analysis. IEEE Transactions on Energy Conversion, 99 (2015) 1-12.
[7] H. Chen, X.L. Zhang, J.L. Duan, X.W. Liu. One-step synthesis of Pt/CeO2-Graphene catalyst by microwave-assisted ethylene glycol process for μDMFC. Materials letters, 126 (2014) 9-12.
[8] X.L. Zhang, Z. Ma, H. Chen, Q. Sun, Y.F. Zhang, X.W. Liu. A micro passive direct methanol fuel cell with high performance via plasma electrolytic oxidation on aluminum-based substrate. Energy, 78 (2014) 149-153.
[9] J.L. Duan, X.L. Liu, H. Chen, Y.F. Zhang. Poly (N-acetylaniline) functionalized graphene nanosheets supported Pt electrocatalysts for methanol oxidation. Microelectronic Engineering, 121 (2014) 100-103.
[10] X.W. Liu, J.L. Duan, H. Chen, Y.F. Zhang, X.L. Zhang. A carbon riveted Pt/Graphene catalyst with high stability for direct methanol fuel cell. Microelectronic Engineering, 110 (2013) 354-357.
[11] Zhang, Yu Feng, et al. "Asymmetric design of anode and cathode current collectors for micro direct methanol fuel cells." Key Engineering Materials. Vol. 645. Trans Tech Publications Ltd, 2015.
[12] Li F, Guo YQ, Liu Y, Yan J, Wang W, Gao JP. Excellent electrocatalytic performance of Pt nanoparticles on reduced graphene oxide nanosheets prepared by a

direct redox reaction between $Na_2PtCl_4$ and graphene oxide. Carbon 2014;67:617-26.
[13] Yuan Z, Zhang Y, Fu W, Li Z, Liu X, Investigation of a small-volume direct methanol fuel cell stack for portable applications. Energy 2013;51:462-7.
[14] Jiang ZZ, Wang ZB, Chu YY, Gu DM, Yin GP, Carbon riveted microcapsule Pt/MWCNTs-$TiO_2$ catalyst prepared by in situ carbonized glucose with ultrahigh stability for proton exchange membrane fuel cell. Energy Environ Sci 2011;4:728-35.
[15] Liu HS, Song CJ, Zhang L, Zhang JJ, Wang HJ, Wilkinson DP. A review of anode catalysis in the direct methanol fuel cell. J Power Sources 2006;155:95-110.
[16] Alayoglu S, Nilekar AU, Mavrikakis M, Eichhorn B. Ru–Pt core–shell nanoparticles for preferential oxidation of carbon monoxide in hydrogen. Nat Mater 2008;7:333-8.
[17] Liu ZL, Ling XY, Su XD, Lee JY. Carbon-supported Pt and PtRu nanoparticles as catalysts for a direct methanol fuel cell. J Phys Chem B 2004;108:8234-40.
[18] Han F, Wang XM, Lian J, Wang YZ. The effect of Sn content on the electrocatalytic properties of Pt–Sn nanoparticles dispersed on graphene nanosheets for the methanol oxidation reaction. Carbon, 2012;50:5498-504.
[19] Lim DH, Lee WD, Choi DH, Kwon HH, Lee HI. The effect of cerium oxide nanoparticles on a Pt/C electrocatalyst synthesized by a continuous two-step process for low-temperature fuel cell. Electrochem Commun 2008;10:592-596.
[20] Zhu JB, Zhao X, Xiao ML, Liang L, Liu CP, Liao JH, Xing W. The construction of nitrogen-doped graphitized carbon–$TiO_2$ composite to improve the electrocatalyst for methanol oxidation. Carbon 2014;72:114-24.
[21] Li QF, He RH, Gao JA, Jensen JO, Bjerrum NJ. The CO poisoning effect in PEMFCs operational at temperatures up to 200 C. J Electrochem Soc 2003;150:A1599-605.
[22] Esparbe I, Brillas E, Centellas F, Garrido JA, Rodriguez RM, Arias C, Cabot PL. Structure and electrocatalytic performance of carbon-supported platinum nanoparticles. J Power Sources 2009;190:201-9.
[23] Zhang ZH, Huang YJ, Ge JJ, Liu CP, Xing W. $WO_3$/C hybrid material as a highly active catalyst support for formic acid electrooxidation. Electrochem Commun 2008;10:1113-6.
[24] Chang Y, Han G, Li M, Gao F. Graphene-modified carbon fiber mats used to improve the activity and stability of Pt catalyst for methanol electrochemical oxidation. Carbon 2011;49:5158-65.
[25] Shao Y, Zhang S, Wang CM, Nie ZM, Liu J, Y Wang, Lin YH. Highly durable graphene nanoplatelets supported Pt nanocatalysts for oxygen reduction. J Power Sources 2010;195:4600-5.
[26] Gu YJ, Wong WT. Nanostructure PtRu/MWNTs as anode catalysts prepared in a vacuum for direct methanol oxidation. Langmuir 2006;22:11447-52.
[27] H. Chen, H. Jia, P.D. Shuvra, J.T. Lin, B.W. Alphenaar, P. X.-L. Feng, "GaN/AlN Heterostructure Micromechanical Self-sustained Oscillator for Middle Ultraviolet (MUV) Light Detection", in Proc. 32nd IEEE International Conference on Micro Electro Mechanical Systems (MEMS 2019), Seoul, Korea, Jan. 27–31, 2019.
[28] H. Chen, H. Jia, V. Pashaei, W. Liao, C.N. Arutt, M.W. McCurdy, P. Hung, R.A. Reed, R.D. Schrimpf, M.L. Alles, P. X.-L. Feng, "Probing Ion Radiation Effects in Si Crystal by 3D Integrated Resonating Thin Diaphragm", AVS 65th International Symposium & Exhibition, Long Beach, CA, Oct. 21-26 (2018).

[29] H. Chen, V. Pashaei, W. Liao, C.N. Arutt, H. Jia, M.W. McCurdy, C.A. Zorman, R.A. Reed, R.D. Schrimpf, M.L. Alles, P. X.-L. Feng, "Radiation Effects on a Silicon Carbide (SiC) Multimode Resonating Diaphragm", 19th IEEE International Conference on Solid-State Sensors, Actuators and Microsystems (Transducers 2017), pp. 990-993, Kaohsiung, Taiwan, Jun. 18-22 (2017).

[30] H. Chen, V. Pashaei, W. Liao, C.N. Arutt, M.W. McCurdy, P. Hung, R.A. Reed, R.D. Schrimpf, M.L. Alles, P. X.-L. Feng, "Ion Radiation Effects in Silicon Carbide (SiC) Crystal Probed by Multimode Diaphragm Resonators", AVS 64th International Symposium & Exhibition, Tampa, FL, Oct. 29-Nov. 3 (2017).

# Figure captions

Fig. 1 Schematic diagram of synthesizing Pt-MnO$_2$/rGO -L catalyst.

Fig. 2 XRD patterns for MnO$_2$, PG, PMG and PMGL.

Fig. 3 Raman spectra of PMGL, PMG and PG catalyst.

Fig. 4 TEM images and corresponding particle size distributions of PMGL(a), PMG(b) and PMGL(c).

Fig. 5 XPS spectra of the Pt 4f region for PMGL, PMG and PG.

Fig. 6 Voltammetry curves for PMGL, PMG and PG in 0.5M H$_2$SO$_4$.

Fig. 7 Voltammetry curves for PMGL, PMG and PG in 0.5M CH$_3$OH + 0.5M H$_2$SO$_4$.

Fig. 8 Chronoamperometry curves for PMGL, PMG and PG in 0.5M CH$_3$OH + 0.5M H$_2$SO$_4$ at 0.65 V.

Fig. 9 Nyquist plot of methanol electrooxidation in 0.5M CH$_3$OH + 0.5M H$_2$SO$_4$.

Fig. 10 Steady-state polarization and power density curves for fuel cells employing PMGL, PMG and PG as anode catalysts.

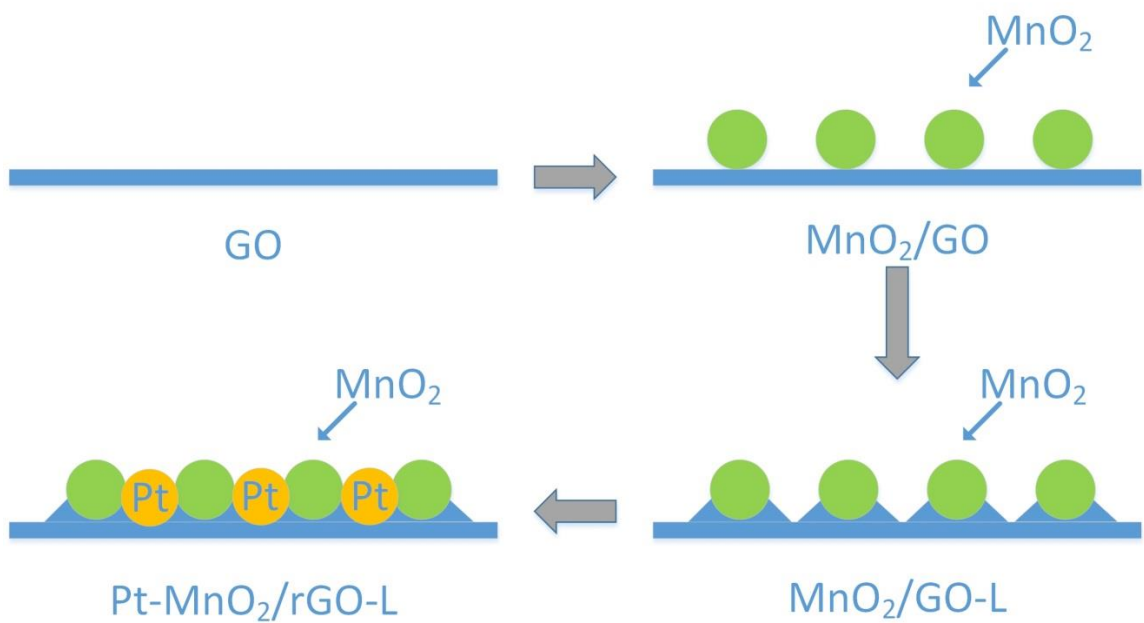

Fig. 1 Schematic diagram of synthesizing Pt-MnO$_2$/rGO-L catalyst.

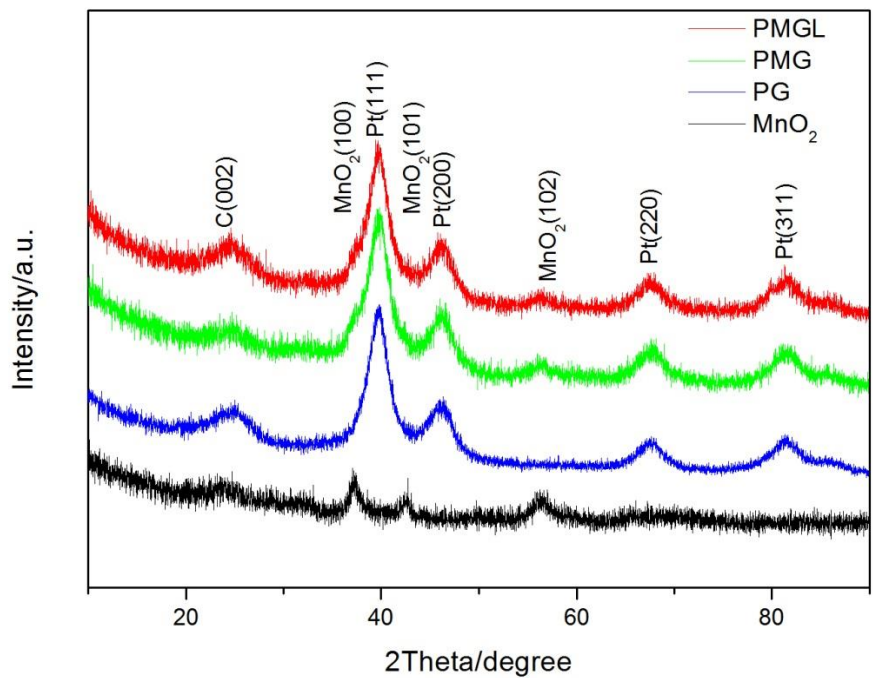

Fig. 2 XRD patterns for $MnO_2$, PG, PMG and PMGL

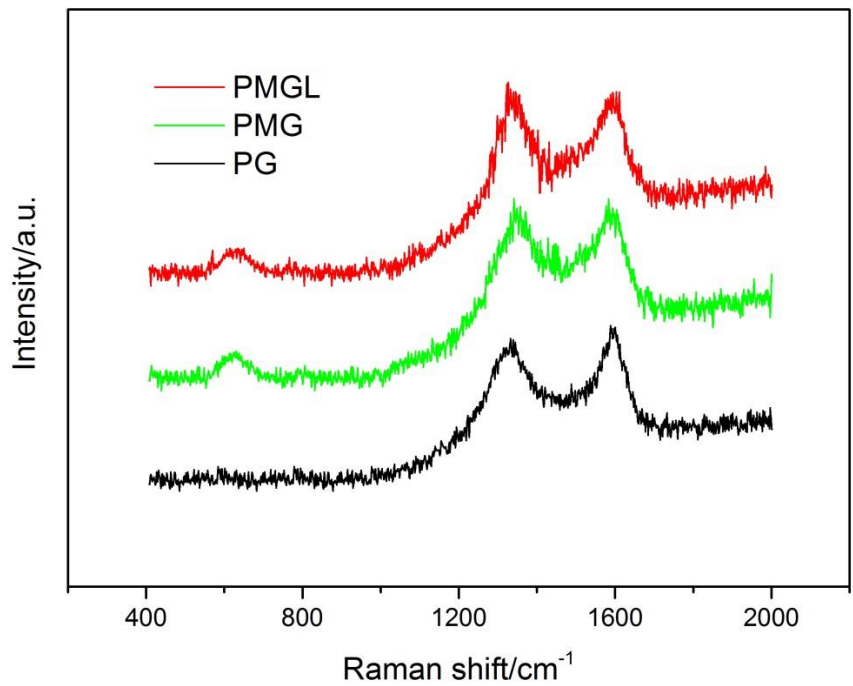

Fig. 3 Raman spectra of PMGL, PMG and PG catalyst.

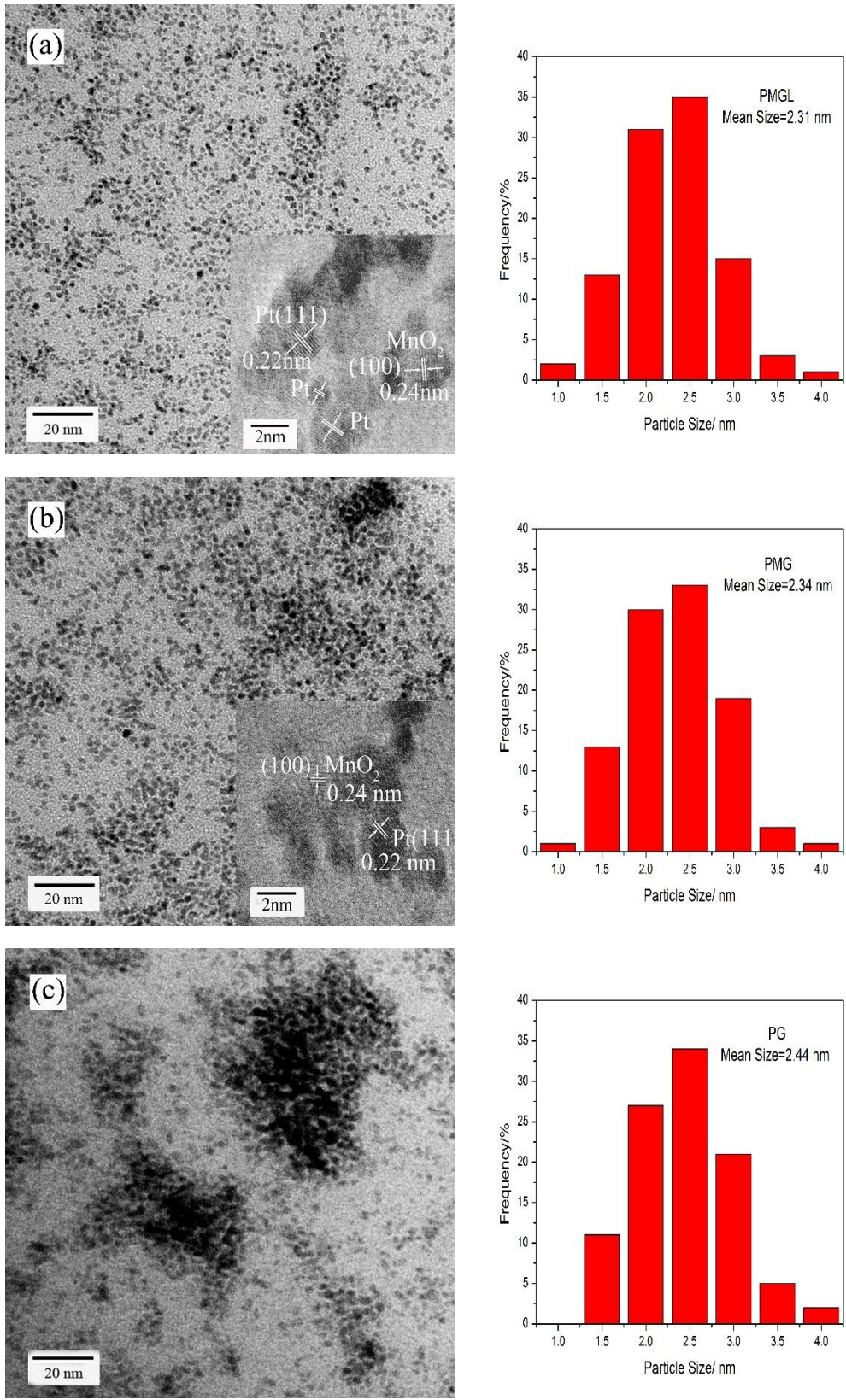

Fig. 4 TEM images and corresponding particle size distributions of PMGL(a), PMG(b) and PMGL(c).

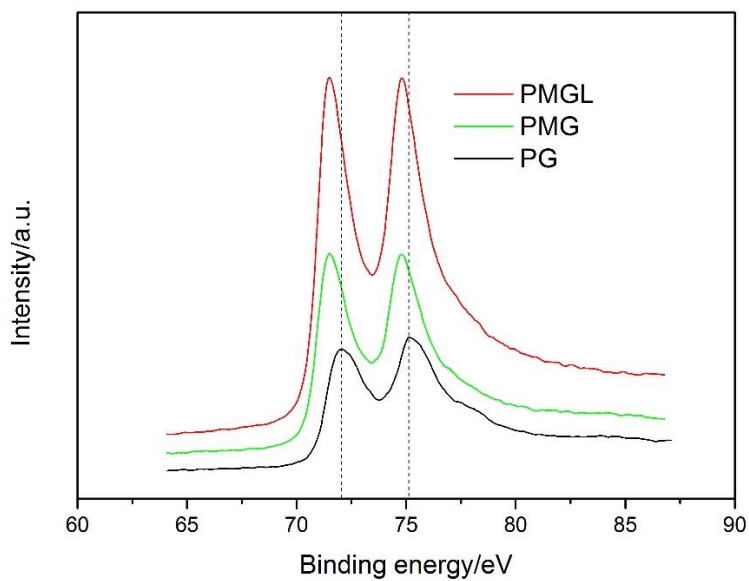

Fig. 5 XPS spectra of the Pt 4f region for PMGL, PMG and PG.

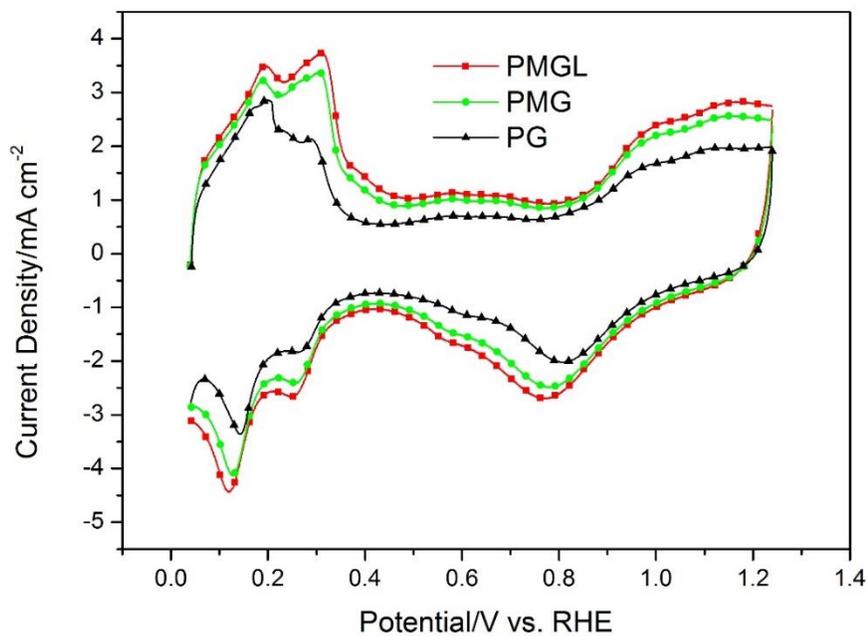

Fig. 6 Voltammetry curves for PMGL, PMG and PG in 0.5M $H_2SO_4$.

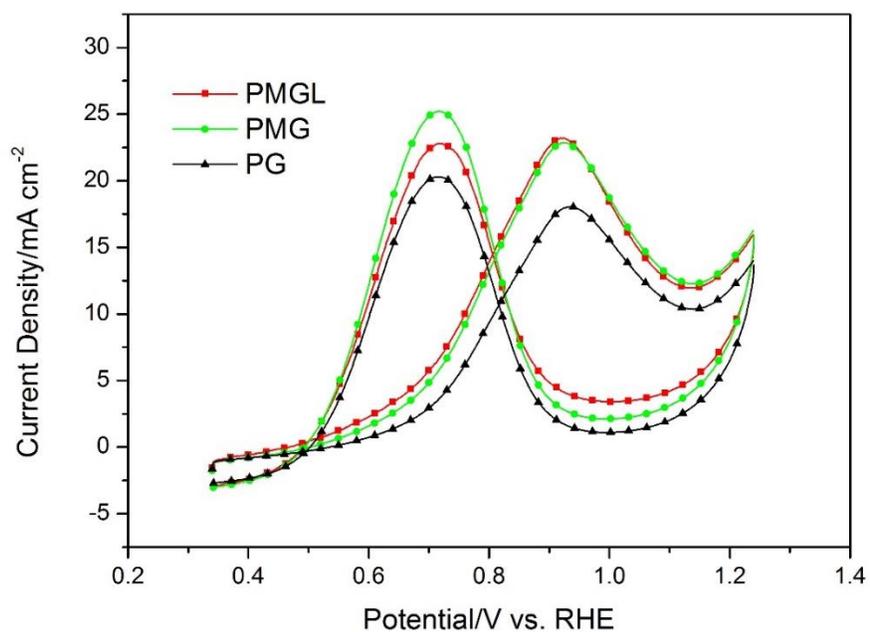

Fig. 7 Voltammetry curves for PMGL, PMG and PG in 0.5M CH$_3$OH + 0.5M H$_2$SO$_4$.

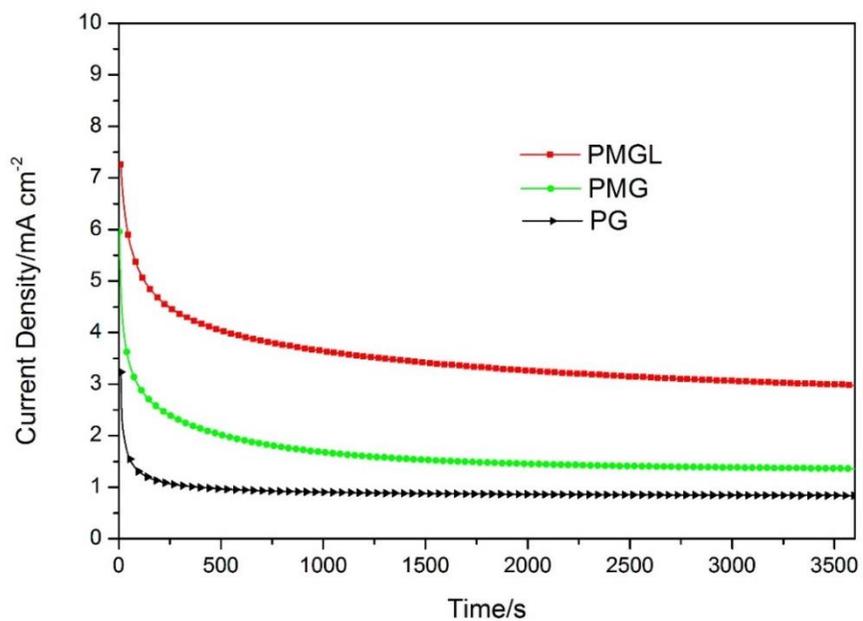

Fig. 8 Chronoamperometry curves for PMGL, PMG and PG in 0.5M CH$_3$OH + 0.5M H$_2$SO$_4$ at 0.65 V.

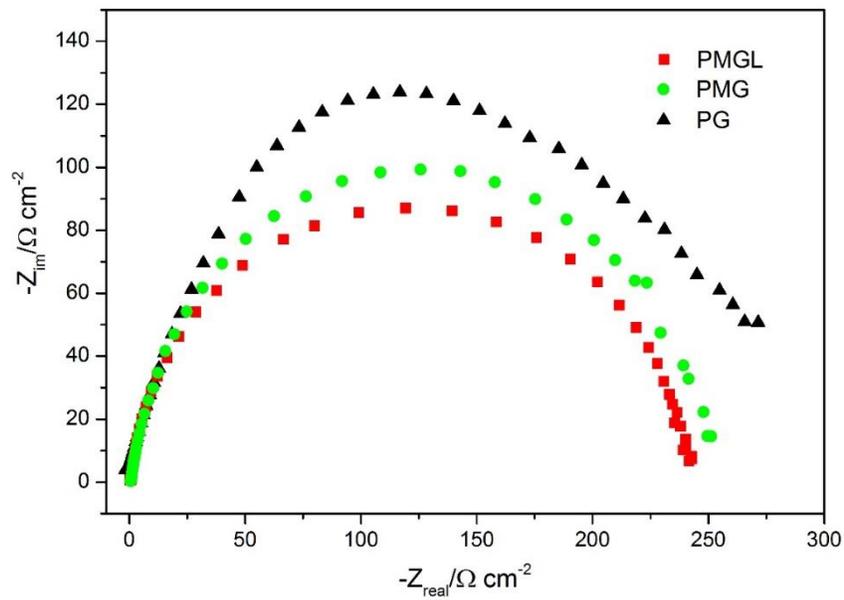

Fig. 9 Nyquist plot of methanol electrooxidation in 0.5M $CH_3OH$ + 0.5M $H_2SO_4$.

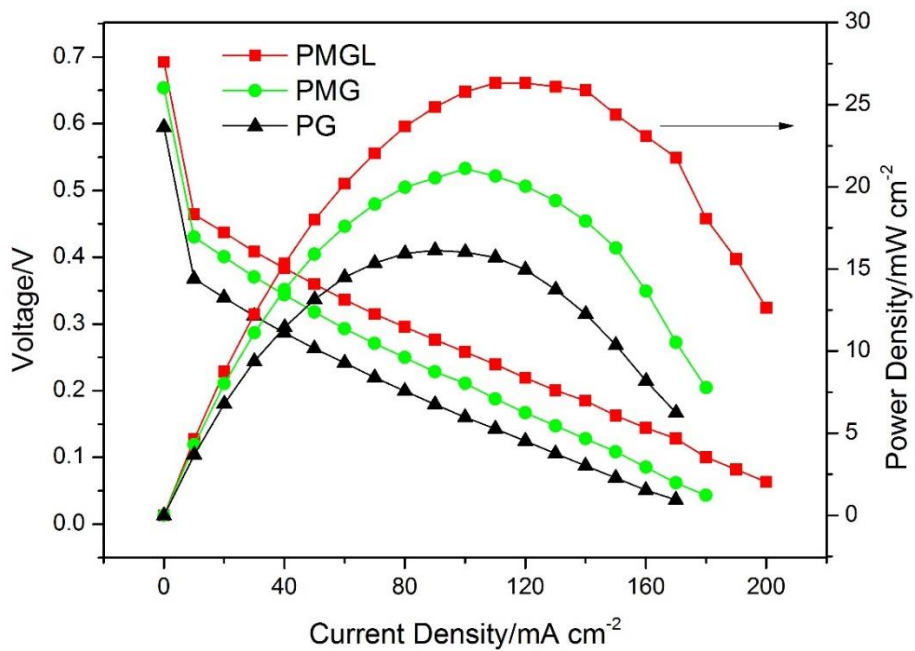

Fig. 10 Steady-state polarization and power density curves for fuel cells employing PMGL, PMG and PG as anode catalysts.